\documentclass[prl,twocolumn,showpacs,preprintnumbers,amsmath,amssymb]{revtex4}

\usepackage{graphicx}
\usepackage{dcolumn}
\usepackage{bm}

\begin{document}

\newcommand{\spinup}{\protect{$ \left|\uparrow \right\rangle$}}
\newcommand{\spindown}{\protect{$ \left|\downarrow \right\rangle$}}
\newcommand{\upeq}{\protect{\left | \uparrow \right\rangle}}
\newcommand{\downeq}{\protect{\left | \downarrow \right\rangle}}
\newcommand{\upxeq}{\protect{\left | \uparrow_x \right\rangle}}
\newcommand{\downxeq}{\protect{\left | \downarrow_x \right\rangle}}

\title{Spin-dependent forces on trapped ions for phase-stable quantum gates and motional Schr\"{o}dinger-cat states}

\author{P.~C. Haljan, K.-A. Brickman, L. Deslauriers, P.~J. Lee, C. Monroe}
\affiliation{FOCUS Center and University of Michigan Department of Physics}

\date{\today}

\begin{abstract}
Favored schemes for trapped-ion quantum logic gates use bichromatic laser
fields to couple internal qubit states with external motion through a
``spin-dependent force." We introduce a new degree of freedom in this
coupling that reduces its sensitivity to phase decoherence. We demonstrate
bichromatic spin-dependent forces on a single trapped $^{111}$Cd$^+$ ion,
and show that phase coherence of the resulting ``Schr\"{o}dinger-cat" states
of motion depends critically upon the spectral arrangement of the optical
fields. This applies directly to the operation of entangling gates on
multiple ions.
\end{abstract}


\maketitle


Trapped atomic ions have a number of desirable features that make them well
suited for quantum information applications \cite{wineland98b}. Pairs of
hyperfine ground states provide an ideal host for quantum bits (qubits) that
can be manipulated and entangled via optical Raman transitions
\cite{blinov2004a}. Since the first proposal for entangling trapped ions
\cite{cirac-zoller95}, significant theoretical advances have relaxed the
constraints on implementation to realize more robust entanglement schemes
\cite{molmer99,solano1999a,milburn2000a,Garcia-Ripoll2003a}. These schemes
rely on a spin-dependent force acting on each ion where ``spin"  refers to
the effective Pauli spin associated with the ion-qubit's two-level system.
Acting on a single ion, the force can entangle the ion's spin degree of
freedom with its motion; following disentanglement, the ion acquires a net
geometric phase that is spin dependent \cite{milburn2000a}. When the force
is applied to two ions, the geometric phase depends nonlinearly on their
spins through their mutual Coulomb interaction, generating entanglement.

Optical Raman fields are a convenient way to create strong spin-dependent
forces for hyperfine ion-qubits \cite{monroe96a,leibfried2003b}. For
example, an entangling gate based on a $\hat{\sigma}_z$-dependent force has
been realized with a state-dependent AC Stark shift from Raman fields
\cite{leibfried2003a}, where $\hat{\sigma}_{x,y,z}$ are the Pauli operators.
Unfortunately, such $\hat{\sigma}_z$ gates are not compatible with magnetic
field insensitive (or ``clock") qubit states
\cite{blinov2004a,hyperfineNIST}, and are therefore open to qubit phase
decoherence from fluctuating magnetic fields. An alternative solution is the
M{\o}lmer-S{\o}rensen (MS) gate \cite{molmer99,sorensen2000a,sackett00},
which uses a more complicated arrangement of bichromatic fields to realize a
$\hat{\sigma}_\phi$-type force where $\hat{\sigma}_\phi$ is a linear
combination of $\hat{\sigma}_x$ and $\hat{\sigma}_y$ operators. While the
gate does work on clock states, it can have a significant phase instability
due to the hyperfine coherences required to generate the spin dependence.
All gates, whatever the spin dependence, are susceptible to fast phase
fluctuations of the Raman fields during the course of gate evolution.
However, the MS gate can also be susceptible to slow phase drifts, which
cause changes in the spin dependence of the force and results in a phase
instability \textit{between} gates.

In this Letter we show how to overcome phase drifts in the MS gate, allowing
clock-state benefits to be realized for ion-based qubits. Such ideas may be
useful in other experimental settings since the MS scheme is a general
paradigm for entanglement \cite{otherMSapps}. A bichromatic force is first
implemented on a single ion to demonstrate the periodic formation of
``Schr\"{o}dinger cat" states of entangled spin and motion \cite{monroe96a}.
The generation of single-ion cat states is a useful test-bed for issues
related to the MS gate without the added complication of two ions. In
particular, we make use of the cat signal to demonstrate a phase-insensitive
gate arrangement.

A $\hat{\sigma}_\phi$ force operates in a spin basis dressed by the Raman
laser fields. It is created from a bichromatic field composed of Raman
couplings to the first vibrational sidebands of the trapped ion's motion,
assumed along the $\hat{z}$-axis. The red sideband coupling
$|\downarrow,n+1\rangle \leftrightarrow |\uparrow,n\rangle$, and the blue
coupling $|\downarrow,n\rangle \leftrightarrow |\uparrow,n+1\rangle$, create
a Jaynes-Cummings and anti-Jaynes-Cummings interaction between the ion's spin
and its vibrational levels $\{|n\rangle\}$\cite{leibfried2003b}. If the red
and blue couplings have the same base Rabi frequencies
($\Omega_r=\Omega_b=\Omega_{sb}$ for $n=0$) and balanced detunings
($-\delta_r=\delta_b=\delta$) then a spin-dependent, oscillating force is
realized under the rotating-wave and Lamb-Dicke approximations
\cite{bichromatic} (see also ref.~\cite{solano2003a}). The Hamiltonian in the
interaction frame is
\begin{equation}\label{eqn:hamiltonian}
H_I(t)= - \frac{F_o z_o}{2}\hat{\sigma}_{\!\phi}(\hat{a}e^{\imath\delta
t+\imath\phi_m}+\hat{a}^\dag e^{-\imath\delta t-\imath\phi_m})
\end{equation}
\noindent where $z_o\!=\!\sqrt{\hbar/2 m \omega_z}$ is the size of the
harmonic oscillator ground state for an ion of mass $m$ and center-of-mass
oscillation frequency $\omega_z$. The strength of the force is given by $F_o
z_o=\hbar\Omega_{sb}$ and $\hat{a}^\dag$ and $\hat{a}$ are the oscillator
raising and lowering operators. The orientation of the force's spin
dependence, $\hat{\sigma}_{\phi}=e^{-\imath\phi_{s}}\hat{\sigma}_+ +
e^{\imath\phi_{s}}\hat{\sigma}_-$, is defined by the azimuthal angle $\phi_s$
where $\hat{\sigma}_\pm$ are the spin raising and lowering operators. Note
$\hat{\sigma}_\phi\!=\!\hat{\sigma}_x$ for the case $\phi_s\!=\!0$. In terms
of the phases $\phi_r$ and $\phi_b$ of the red and blue sideband fields, the
spin phase of the MS force in Eqn.~\ref{eqn:hamiltonian} is
$\phi_{s}\!=\!(\phi_b+\phi_r)/2$ and the motional phase is
$\phi_{m}\!=\!(\phi_b-\phi_r)/2$. The unitary evolution generated by the
force is a spin-dependent displacement $\hat{\mathcal{U}}(t)=
\hat{\mathcal{D}}(\alpha(t))
\protect{\left|\uparrow_\phi\rangle\langle\uparrow_\phi\right|} +
\hat{\mathcal{D}}(-\alpha(t))\protect{\left|\downarrow_\phi\rangle\langle\downarrow_\phi\right|}$
where $\protect{\left|\downarrow_\phi\right\rangle}$ and
$\protect{\left|\uparrow_\phi\right\rangle}$ are the eigenstates of
$\hat{\sigma}_\phi$ \cite{geom-phase}. $\hat{\mathcal{D}}(\alpha)$  is the
displacement operator in position-momentum ($z$-$p$) phase space with
$\alpha(t)=\alpha_o e^{-i\phi_m}(1-e^{-i\delta t})$ and
$\alpha_o=\Omega_{sb}/2\delta$. The corresponding position and momentum
displacements are $2 z_{o} Re(\alpha(t))$ and $2m\omega_{z} z_{o}
Im(\alpha(t))$ respectively.

\begin{figure}
\begin{center}
\includegraphics[width=\linewidth,clip]{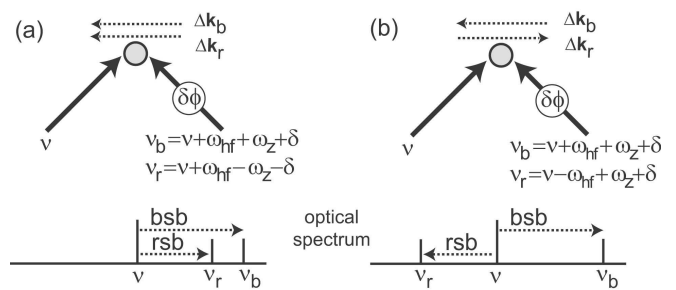}
\end{center}
\caption{Two possible Raman setups (top) to realize a bichromatic force on a
hyperfine ion-qubit and corresponding optical spectrum (bottom) showing red
(rsb) and blue (bsb) sideband beatnotes. Two laser beams (solid lines, top)
with optical carrier frequency $\nu$ generate (a) co-propagating or (b)
counter-propagating Raman running waves (dotted lines) at the ion with
wavevectors $\Delta\textbf{k}_{r,b}$. Symbols include optical phase
fluctuation $\delta\phi$, hyperfine splitting $\omega_{hf}$, ion trap
frequency $\omega_z$, and force detuning $\delta$.} \label{fig:Ramansetup}
\end{figure}

Instabilities in the spin and motional phases $\phi_s$ and $\phi_m$ arise
from optical phase fluctuations $\delta\phi$ between the laser beams
generating the Raman field at the ion [Fig.~\ref{fig:Ramansetup}], or
equivalently from fluctuations in the ion center-of-mass position. Slow
global fluctuations in the force's motional phase are unimportant in the
context of MS entangling gates, which involve only transient motional
coherences associated with closed trajectories in phase space. Motional
coherence between gates is therefore irrelevant. Such is not the case for
spin coherences; however, a bichromatic field offers an additional degree of
freedom to suppress the effect of fluctuations on $\phi_s$. The two fields
can be composed of co- or counter- propagating Raman beat waves. For the
co-propagating (phase sensitive) setup in Fig.~\ref{fig:Ramansetup}(a), the
phase fluctuation $\delta\phi$ appears on the red and blue sideband
couplings with the same sign ($\phi_r=\phi_b=\delta\phi$) and so is directly
written onto the MS spin phase ($\phi_s=\delta\phi$). In the
counter-propagating (phase insensitive) case in
Fig.~\ref{fig:Ramansetup}(b), the phase fluctuation appears with the
opposite sign ($-\phi_r=\phi_b=\delta\phi$) and is thus eliminated from the
spin phase ($\phi_s\equiv0$). The phase-sensitive setup can be useful for
cancelling common-mode fluctuations with other gate operations employing the
same Raman beams while the phase-insensitive setup is useful for
synchronizing with other operations involving microwave fields or different
Raman beam geometries. This discussion only accounts for fluctuations shared
by the red and blue sideband fields as shown in Fig.~\ref{fig:Ramansetup},
but independent fluctuations can also be avoided with an appropriate optical
design.

Our experiment is centered around a linear ion-trap \cite{Deslauriers2004a}
employing both rf and dc fields to create a harmonic confining potential,
characterized by trapping frequencies
$\{\omega_x,\omega_y,\omega_z\}/2\pi=\{8,9,3.55\}$MHz. Single $^{111}$Cd$^+$
ions are loaded into the trap from a neutral CdO oven source via resonant
pulsed-laser photoionization. The qubit resides in the $5S_{1/2}$ ground
hyperfine clock states, $\downeq=\left|F=1,m_F=0\right\rangle$ and
$\upeq=\left|F=0,m_F=0\right\rangle$, separated by
$\omega_{hf}/2\pi=14.53$GHz. Doppler cooling on the $5S_{1/2}-5P_{3/2}$
cycling transition (wavelength 214.5nm, linewidth $\gamma/2\pi=47$MHz) brings
the ion to a thermal state given by $\bar{n}\approx6$. An optional step of
sideband cooling can further cool the ion to near the ground state of motion
along the $\hat{z}$-axis ($\bar{n}\approx0.05$) \cite{Deslauriers2004a},
well within the Lamb-Dicke limit. Optical pumping allows near perfect
preparation of the qubit to the state \spinup. High efficiency qubit
detection is accomplished through fluorescence detection, where strong ion
fluorescence is indicative of state \spinup~whereas state \spindown~remains
nearly dark.

Raman transitions are generated with a pair of laser beams in a 90$^\circ$
geometry with wavevector difference $\Delta$\textbf{k} along the
$\hat{z}$-axis [Fig.~\ref{fig:Ramansetup}]. Coupling occurs via the
$P_{3/2}$ state from which the beams are detuned by $\sim$220GHz. A
combination of laser modulation techniques is used to generate the Raman
beat notes \cite{lee2003a}. To bridge the hyperfine gap an electro-optic
modulator is used to impart a microwave frequency comb onto the Raman laser,
which is then split into two arms. Since the comb exists on both arms, all
frequencies exist to access beat waves propagating in either direction. A
pair of acousto-optic modulators (AOMs), one in each arm, generates a
frequency offset between the two beams. This allows for tuning across the
various motional sideband resonances associated with either wavevector
direction. A bichromatic Raman field is obtained by driving one AOM with two
rf frequencies [Fig.~\ref{fig:Ramansetup}] so that any optical phase
fluctuations on the red and blue sidebands are common mode. The sideband Rabi
frequencies, typically 2kHz, are balanced \textit{in situ} to better than
10\% and the detunings $\delta_r$ and $\delta_b$ to better than 100Hz.

The action of the bichromatic force is demonstrated by applying it to a
single ion, initially prepared in \spinup. This initial state can be
expressed in terms of the diagonal spin basis of the bichromatic force as
$\frac{1}{\sqrt{2}}(\upxeq+\downxeq)|0\rangle$, where without loss of
generality the force is taken to be $\hat{\sigma}_x$-dependent and the
initial motional state is the ground state. Application of the force
generates the entangled state $\frac{1}{\sqrt{2}}(\upxeq |\alpha\rangle +
\downxeq\left|-\alpha\right\rangle)$, which can be written in the measurement
or $\hat{\sigma}_z$ basis as $\upeq
\left(|\alpha\rangle+\left|-\alpha\right\rangle\right)/2 +
\downeq\left(|\alpha\rangle-\left|-\alpha\right\rangle\right)/2$
\cite{monroe96a,deMatosFilho1996a}. Taking into account an initial thermal
motional state as well as weak decoherence from motional heating
\cite{turchette2000a} leads to the general expression for the probability of
measuring the ion to be in state \spindown:
\begin{equation}\label{eqn:cat}
P_{\downarrow}^c(\tau)=\frac{1}{2}[1- e^{ -
\frac{1}{2}\dot{\bar{n}}\tau|4\alpha_o|^2-(\bar{n}+\frac{1}{2})|2
\alpha(\tau)|^2}]
\end{equation}
\noindent The initial thermal distribution is characterized by average
occupation number $\bar{n}$, the motional heating rate by $\dot{\bar{n}}$
and the duration of the force by $\tau$.
\begin{figure}
\begin{center}
\includegraphics[width=\linewidth,clip]{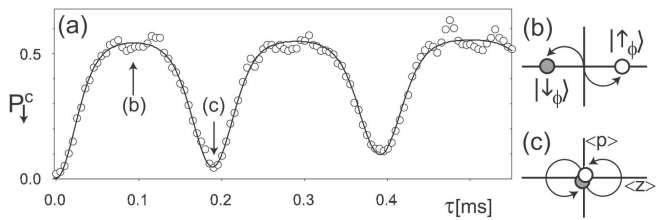}
\end{center}
\caption{Single-ion evolution due to a spin-dependent bichromatic force. (a)
Probability $P^c_\downarrow$ of measuring state \spindown ~plotted vs. force
duration $\tau$. Ion is initially Doppler cooled. Data is run-time averaged
(100 shots/point) and smoothed. Solid line is a fit to Eqn.~\ref{eqn:cat},
modified to include a linear change in peak and contrast (from spontaneous
emission) and a detuning drift across the data. Fit parameters include
$\Omega_{sb}/2\pi$=2.2kHz(fixed), $\delta/2\pi$=--5.46(3)kHz and
$\bar{n}$=8.1(3). (b)\&(c)Phase space sketches of the ion motion at points
indicated in (a).} \label{fig:timescan}
\end{figure}

An experiment in which the duration of the force is varied before qubit
detection demonstrates the periodic entanglement in single ion evolution
[Fig.~\ref{fig:timescan}]. A transition to $P_{\downarrow}^c=1/2$ indicates
when a Schr\"{o}dinger cat is formed. At this point, the motional wavepackets
of the cat state are sufficiently far separated that the spin interference
is inhibited, yielding equal probability of \spinup ~and \spindown ~in the
measurement basis. At fringe minima corresponding to $\delta\tau=2m\pi$ with
$m$ an integer, the motional states return to their original position and
overlap. The spin interference is restored, giving the initial state
\spindown ~such that $P_\downarrow^c\approx0$. This coherent process of
periodic entanglement and disentanglement of the spin and motional degrees
of freedom continues with reasonable signal quality for at least two
oscillations [Fig.~\ref{fig:timescan}].

\begin{figure}
\begin{center}
\includegraphics[width=\linewidth,clip]{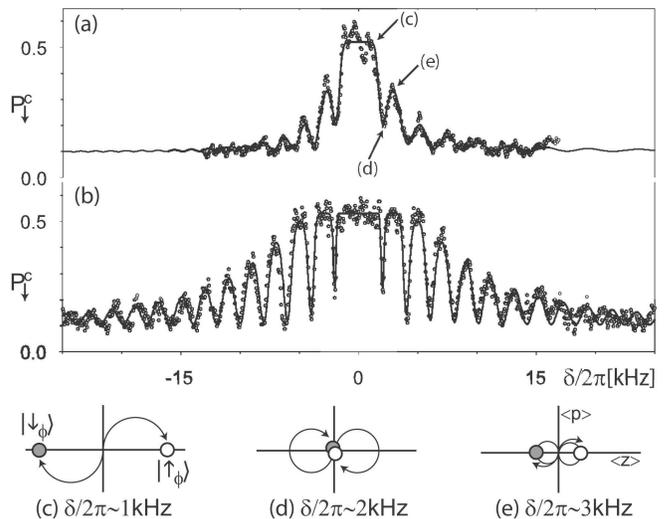}
\end{center}
\caption{Probability $P^c_\downarrow$ plotted vs. detuning $\delta$ of the
bichromatic force for (a) a ground-state cooled and (b) a Doppler cooled
ion, initially prepared in \spinup. The force is applied for 500$\mu$s. Data
is run-time averaged with 100 shots/point and smoothed. Solid lines show fits
to \protect{Eqn.~\ref{eqn:cat}} modified to include overall peak and
contrast factors (for spontaneous emission) and a detuning drift across the
data. An initial fit for data in (a) with $\bar{n}$=0.05 fixed gives
$\Omega_{sb}/2\pi$=1.62(3)kHz and $\dot{\bar{n}}$=0.44(2)ms$^{-1}$. A
subsequent fit for data in (b) with $\Omega_{sb}/2\pi$=1.62kHz fixed gives
$\bar{n}$=5.6(1) and $\dot{\bar{n}}$=0.62(6)ms$^{-1}$. The values of
$\dot{\bar{n}}$ are 2-3 times larger than the directly measured heating rate
0.2ms$^{-1}$ \protect{\cite{Deslauriers2004a}}. Phase space sketches
(c)--(e) indicate ion evolution at detunings referenced in (a).}
\label{fig:freqscan}
\end{figure}

A complementary experiment in which the duration of the force is fixed but
the detuning $\delta$ is varied is shown in Fig~\ref{fig:freqscan}. We have
performed this experiment for two different initial temperatures of the ion,
a ``hot" case of a Doppler cooled ion and a ``cold" case of a ion cooled to
near the ground state. On resonance where the force is strongest, the
inferred cat state separation in Fig.~\ref{fig:freqscan} is $\Delta z\approx
10z_o$, a factor of 2.8 larger than the rms size of the ``hot" ion's thermal
state. The hot case has a broader envelope and narrower fringes. This occurs
because the average over the initial thermal distribution quickly draws the
experiment outcome towards $P_\downarrow^c=1/2$, even for small
displacements. Nevertheless, all initial states should return on themselves
at the same moment (within the Lamb-Dicke approximation) giving a full
revival. The overall decrease of contrast, particularly visible in the
non-zero baseline, is due to spin decoherence and optical pumping induced by
spontaneous emission. The detuning-dependent fringe contrast is consistent
with motional decoherence which has a characteristic exponential sensitivity
to the size of the cat state, largest near resonance ($\delta\sim0$)
\cite{walls85,turchette2000a}.

In order to observe the phase sensitivity of the MS force an interference
technique is required. Ramsey interferometry combined with Schr\"{o}dinger
cat formation provides a fluorescence signal that is sensitive to the
orientation of the force's spin basis, characterized by the phase $\phi_s$.
The essence of the procedure is to rotate the spin of the initial state
$\upeq$ with a $\pi/2$ pulse into the ``$x$-$y$" plane at some azimuthal
angle $\phi_o$. This results in the state
$1/\sqrt{2}(\upeq+e^{\imath\phi_o}\downeq)$ where $\phi_o$ is the reference
phase associated with the $\pi/2$ pulse. The MS force is then applied.
Finally a 3$\pi$/2 analysis pulse returns the spin to the ``$z$-axis" before
measurement. A signal sensitive to $\phi_s$ is obtained as follows:
\begin{equation}\label{eqn:ramsey}
P_\downarrow=P_\downarrow^c(\tau)\sin^2(\phi_o-\phi_s)
\end{equation}
\noindent where $P_{\downarrow}^{c}(\tau)$ is given by Eqn.~\ref{eqn:cat}. As
long as the detuning and duration $\tau$ of the force are chosen to generate
a significant displacement ($\alpha\gg1$), the signal is approximately
$\frac{1}{2}\sin^2(\phi_o-\phi_s)$. If the initial $\pi/2$ pulse rotates the
ion's spin into a state in which the force is diagonal ($\phi_o=\phi_s+m\pi$
with $m$ an integer), a displacement occurs but no entangled cat state is
formed. Therefore following the analysis pulse, the ion's spin returns to its
initial state \spinup ~such that $P_\downarrow=0$. On the other hand, if the
rotated initial state deviates from this special condition
($\phi_o\neq\phi_s+m\pi$), then the state is a superposition of the force's
spin basis and a Schr\"{o}dinger cat is formed. This is revealed by a net
fluorescence signal on analysis.
\begin{figure}
\begin{center}
\includegraphics[width=\linewidth,clip]{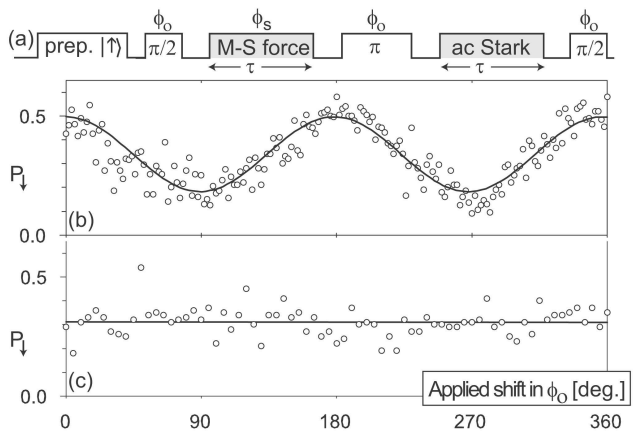}
\end{center}
\caption{(a) Interferometric photon-echo sequence to test optical phase
sensitivity of the MS force. Phase $\phi_o$ and duration of spin-rotation
pulses (unshaded) indicated. The MS pulse (with spin phase $\phi_s$ ) and
pulse for ac Stark shift compensation are shown shaded. (b)\&(c) Probability
$P_\downarrow$ plotted vs. applied shift in $\phi_o$ where (b) and (c)
correspond to setups in \protect{Figs.~\ref{fig:Ramansetup}}(a)\&(b)
respectively. Data is run-time averaged with 100 shots/point requiring
$\sim$200ms/point. Parameters include $\pi/2$ pulse time of 13$\mu$s,
$\Omega_{sb}/2\pi$$\approx$2kHz, $\delta/2\pi$$\approx$5kHz, $\tau$=90$\mu$s,
and $\bar{n}$$\approx$6. Solid lines are a sinusoidal fit (b) and linear fit
(c).} \label{fig:phasescan}
\end{figure}

For convenience, we make use of an optical Raman carrier transition to drive
the $\pi/2$ pulses. The actual pulse sequence for the interferometry employs
a photon echo scheme where the the $3\pi/2$ analysis pulse is divided into a
$\pi$ and $\pi/2$ pulse [Fig.~\ref{fig:phasescan}(a)]. This provides a
convenient way to cancel the effect of Stark shifts ($\sim$20kHz) from the
optical carrier pulses as well as from the MS force itself. The
Stark-induced phase from the MS pulse is cancelled by introducing a
bichromatic pulse, far detuned from the motional resonance, into the second
echo zone [Fig.~\ref{fig:phasescan}(a)].

The carrier pulses which act as a phase reference are subject to the same
optical phase drifts as the sidebands creating the bichromatic force.
Therefore the interferometric signal $P_\downarrow$ is expected to be stable
for the phase sensitive Raman setup [Fig.~\ref{fig:Ramansetup}(a)] where
fluctuations in the force's spin phase $\phi_s$ are common to the reference
phase $\phi_o$. A scan of the reference phase [Fig.~\ref{fig:phasescan}(b)]
shows interference fringes with nearly full contrast (maximum being 1/2).
This remains the case even with a piezoelectric transducer slowly (1Hz)
modulating one Raman arm by an optical wavelength. Wash-out of the
interferometric signal should occur when the MS force is generated with the
phase insensitive setup [Fig.~\ref{fig:Ramansetup}(b)]. In this case, the
force's spin dependence, now decoupled from instabilities, is no longer
correlated with the optical phase fluctuations on the carrier pulses. This
configuration did not require the addition of any Raman path-length
modulation to spoil the fringe contrast as inherent phase drifts on the
optical table were a sufficient source of noise over the 200ms experiment
averaging time.

A spin-dependent force for clock-state qubits has been demonstrated that can
be combined with other operations in a way that is insensitive to slow
optical phase fluctuations in the Raman beams and to drifts in ion
center-of-mass position. This opens the way to phase-stable algorithms as
follows: Single qubit rotations can be implemented with a co-propagating
Raman transition or microwaves. The entangling gate may be realized directly
with a $\hat{\sigma}_\phi$-dependent force based on the phase-insensitive
Raman setup. Alternatively, the phase-sensitive setup can be used in an echo
sequence as in Fig.~\ref{fig:phasescan}(a) to construct a $\hat{\sigma}_z$
gate piecewise where the carrier pulses act as a
$\hat{\sigma}_z\leftrightarrow\hat{\sigma}_x$ basis transformation. This
arrangement can also suppress phase noise due to slow fluctuations in the
relative position of two ions.

This work was supported by the U.S. National Security Agency and Advanced
Research and Development Activity under Army Research Office contract
DAAD19-01-1-0667 and the National Science Foundation Information Technology
Research Program.

\bibliographystyle{apsrevmod}

\begin{thebibliography}{22}
\bibitem[{geo()}]{geom-phase}
\bibinfo{note}{A global displacement-induced phase $\theta(t)$ has been omitted
  from $\hat{U}(t)$.}
\end{thebibliography}


\end{document}